\documentclass[preprint2]{aastex}

\usepackage{amsmath}
\usepackage{import}
\usepackage{graphicx}
\usepackage{subfigure}
\usepackage{dblfloatfix}
\usepackage{threeparttable} 
\usepackage{booktabs} 
\usepackage{array} 

\usepackage{pst-node}

\usepackage{textcmds}
\usepackage[bookmarks]{hyperref} 
\usepackage{natbib}
\bibpunct{(}{)}{;}{a}{}{,}

\usepackage{color} 

\newcolumntype{L}[1]{>{\raggedright\arraybackslash}p{#1}} 
\newcolumntype{C}[1]{>{\centering\arraybackslash}m{#1}} 
\newcolumntype{R}[1]{>{\raggedleft\arraybackslash}p{#1}} 

\shorttitle{Photophoretic Strength on Chondrules 2: Experiment}
\shortauthors{Loesche et al.}

\begin{document}

   \title{PHOTOPHORETIC STRENGTH ON CHONDRULES. 2. EXPERIMENT}

   \author{
          Christoph Loesche,
          Jens Teiser,
          Gerhard Wurm,
          Alexander Hesse
          }
   \affil{Faculty of physics, University of Duisburg-Essen, Lotharstr. 1, D-47057 Duisburg, Germany}
   \email{christoph.loesche@uni-due.de}
   
   \author{Jon M. Friedrich}
   \affil{Department of Chemistry, Fordham University, Bronx, NY 10458
              and Department of Earth and Planetary Sciences, American Museum of Natural History, New York, NY 10024}
   
   \author{Addi Bischoff}
   \affil{Institut f\"ur Planetologie, Westf\"alische Wilhelms-Universit\"at M\"unster, Wilhelm-Klemm-Str. 10, D-48149 M\"unster, Germany}

   \date{ Received \today; accepted }
   
   \begin{abstract}
      Photophoretic motion can transport illuminated particles in protoplanetary disks. In a previous paper we focused on the modeling of steady state photophoretic forces based on the compositions derived from tomography and heat transfer. Here, we present microgravity experiments which deviate significantly from the steady state calculations of the first paper. The experiments on average show a significantly smaller force than predicted with a large variation in absolute photophoretic force and in the direction of motion with respect to the illumination. Time-dependent modeling of photophoretic forces for heat-up and rotation show that the variations in strength and direction observed can be well explained by the particle reorientation in the limited experiment time of a drop tower experiment. In protoplanetary disks, random rotation subsides due to gas friction on short timescales and the results of our earlier paper hold. Rotation has a significant influence in short duration laboratory studies. Observing particle motion and rotation under the influence of photophoresis can be considered as a basic laboratory analog experiment to Yarkovsky and YORP effects.
   \end{abstract}


\section{INTRODUCTION}\label{sec:Introduction}

This paper complements work published by \citet{Loesche2013} (hereafter Paper I) on the photophoretic motion of chondrules. In our first paper, the role of photophoresis was calculated in the following way. (1) Tomography was used to deduce the composition and the geometry of a set of chondrules. (2) Heat transfer through these particles was calculated with illumination heating the chondrule from one side. (3) The photophoretic force was calculated from the resulting temperature distribution along the surface. 

In this paper, we report on experiments at the drop tower of the Center of Applied Space Technology and Microgravity at the University of Bremen where we measured the photophoretic force on the same chondrules that were tomographed and modeled in Paper I, plus a small number of additional chondrules from the same meteorite not tomographed and modeled. As outlined below, we also extend the numerical modeling of heat transfer to rotating chondrules.

The basics of photophoresis are given in Paper I. In short, photophoresis is a possible transport mechanism in protoplanetary disks. The idea was introduced to the field by \citet{Krauss2005} and \citet{Wurm2006}. If particles are embedded in a gaseous environment and are illuminated with a directed radiation, they experience a force, which is usually directed away from the light source \citep{Cheremsin2005}. Times and locations for photophoresis to work can be 
\begin{itemize}
\item at a later time within optically thin protoplanetary disks as a whole \citep{Mousis2007,Krauss2007,Moudens2011,Takeuchi2008,Herrmann2007,Borstel2012},
\item the optical surface of protoplanetary disks \citep{Mousis2007, Wurm2009, Kelling2009,Eymeren2012}, 
\item and the inner edge of protoplanetary disks \citep{Loesche2012, Wurm2009,Beule2013,Kelling2011,Wurm2007,Haack2007,Kelling2013}.
\end{itemize}
Applications include the general transport, e.g., of hot minerals to the outer regions, or potentially the size sorting of chondrules as observed in different chondrites that are otherwise compositionally related.

While some initial ideas on particle transport and sorting have been proposed in the papers mentioned above, the existing description of photophoresis was not accurate enough to consider the details in dedicated transport models.
This is seen in measurements of the photophoretic force on chondrules by \citet{Wurm2010}, which showed strong variations in the photophoretic force of uncertain origin.
The photophoretic force on spherical particles has been quantified by a number of authors \citep{Rohatschek1995,Beresnev1993,Cheremsin2005}.
A number of experiments on photophoresis have also been carried out \citep{Ehrenhaft1918,Rohatschek1956a,Rohatschek1956b,Steinbach2004,Borstel2012}.
Theoretical descriptions and experiments generally agree to within an order of magnitude.
Deviations between theoretical predictions and experiment are often attributed to uncertain parameters such as shape, optical or thermal properties (thermal conductivity), and theoretical approximations.
However, to thoroughly treat particle sorting, a more accurate treatment of photophoresis is essential.

\begin{figure}[h!]
   \centering
   \includegraphics[width=.66\columnwidth]{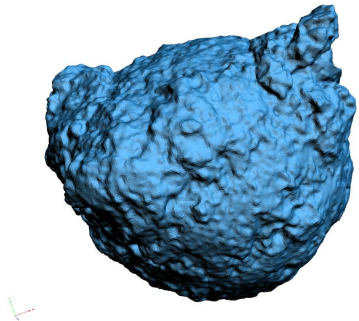}
   \caption{Example of a chondrule having a not-perfect spherical shape, but for which the average photophoretic force can be calculated assuming a spherical particle with the same volume.}
   \label{fig:Particle}
\end{figure}

In view of the factor of three deviations between descriptions and numerical modeling of photophoresis, \citet{Loesche2012} considered the photophoretic force in the molecular flow regime for spherical particles (dust mantled and bare silica spheres) in more detail.
They presented a new analytical equation for spherical particles that is highly accurate within a few percent.
Paper I then calculated the photophoretic force on almost but not perfect spherical chondrules (Figure \ref{fig:Particle}) with heterogeneous composition.

We found that samples from the LL4 chondrite Bjurb\"ole could be well described by a spherical particle using (Paper I):
\begin{equation}
\begin{split}
 F &= F\left(r,k,\alpha,I,T_\mathrm{gas}^\mathrm{kin}, T_\mathrm{gas}^\mathrm{opt} \right)\\
   &= \left(0.7231-0.1741 e^{-2.180 \frac{r}{k}} + 0.4316 e^{-0.9251 \alpha} \right) *\\
   & \quad * \alpha \, \frac{\pi}{6} \, \frac{p}{T_\mathrm{gas}^\mathrm{kin}} \, I \, r^2  \left[ \frac{k}{r}+4\sigma\left(\frac{I}{4\sigma} + \left( T_\mathrm{gas}^\mathrm{opt} \right)^4\right)^{\frac{3}{4}} \right]^{-1}
 \; , \label{eq:forceNumerical}
\end{split}
\end{equation}
where $p$ is the gas pressure, $I$ is the light flux, and $T_\mathrm{gas}^\mathrm{opt}$ and $T_\mathrm{gas}^\mathrm{kin}$ are the temperatures of the radiation field and the temperature of the gas, respectively.
The latter two can be different in optical thin environments. The accommodation coefficient $\alpha$ and, more important, the particle radius $r$ and the thermal conductivity $k$ enter as particle properties. We found that chondrules can be described by a radius equivalent to the radius of a sphere of the same volume. The thermal conductivity is determined by the ratio of the two most abundant phases in the chondrule, silicates and fine-grained material. If we attribute a thermal conductivity of 4.6 to the silicates and 0.1 to the fine-grained material, the total thermal conductivity of a chondrule can be described as (Paper I) 
\begin{equation}
	k = -1.87\cdot 10^{-7} x^3 + 2.94\cdot 10^{-4} x^2 - 7.22\cdot 10^{-2} x + 4.54 \; , \label{eq:kappaChondrule}
\end{equation}
where $x$ is the fine-grained devitrified mesostasis fraction in the corresponding two-phase system.
Some basic parameters of the chondrules deduced in Paper I and used in the experiments are shown in Table \ref{tab:results}.

\begin{table*}
\scriptsize
\centering
\caption{Properties of Chondrules Used Both in the Experiment and for Numerical Studies.}

\begin{threeparttable}
\begin{tabular}{>{\bfseries}C{.4cm} C{.7cm} C{.6cm} C{.8cm} C{.6cm} C{.6cm} C{.5cm} C{.5cm} >{\bfseries}C{.5cm} C{.4cm} >{\bfseries}C{.5cm} C{1.1cm}} \toprule
\multicolumn{2}{c}{} & \multicolumn{5}{c}{\textbf{Compounds} (vol \%)} & \multicolumn{3}{c}{\textbf{Radii} $(\mathrm{mm})$} & \multicolumn{2}{c}{\textbf{Eff. Thermal}} \\
\multicolumn{7}{c}{} & \multicolumn{3}{c}{\tiny (Ref. Point: Center of Shape)} & \multicolumn{2}{c}{\textbf{Conductivity} $\boldsymbol{k} \left( \mathrm{\frac{W}{m\cdot K}}\right)$} \\
\cmidrule (l){3-7} \cmidrule (l){8-10} \cmidrule (l){11-12}
 Sample & Total\newline Volume ($\mathrm{mm^3}$) & Porosity & Fine-\newline Grained\newline Areas\newline (Feldspar-Normative) & Olivines and \newline Pyroxenes & FeS (Troilite) & Fe,Ni-metal & $r_{\max}$ & $\boldmath r_{\text{s}}$ & $r_{\min}$ & Mean & STD Interval \\ 
\midrule
 \text{1} & 5.065 & 1.4 & 21.4 & 74.6 & 2.0 & 0.6 & 1.36 & 1.065 & 0.82 & 2.60 & 2.54 -- 2.67 \\
 \text{2} & 0.020 & 2.3 & 38.0 & 59.6 & 0.0 & 0.0 & 0.24 & 0.167 & 0.09 & 2.28 & 2.06 -- 2.57 \\
 \text{3} & 0.096 & 3.0 & 34.5 & 62.5 & 0.0 & 0.0 & 0.36 & 0.284 & 0.20 & 1.84 & 1.79 -- 1.90 \\
 \text{4} & 0.398 & 2.1 & 31.0 & 65.6 & 1.2 & 0.2 & 0.56 & 0.456 & 0.34 & 2.06 & 2.00 -- 2.13 \\
 \text{5} & 0.369 & 0.6 & 27.4 & 70.7 & 1.3 & 0.0 & 0.53 & 0.445 & 0.38 & 3.11 & 3.04 -- 3.19 \\
 \text{6} & 0.227 & 0.1 & 22.5 & 77.3 & 0.0 & 0.0 & 0.43 & 0.379 & 0.28 & 3.51 & 3.47 -- 3.54 \\
 \text{7} & 0.189 & 4.3 & 39.4 & 55.3 & 1.0 & 0.1 & 0.41 & 0.356 & 0.25 & 2.19 & 2.13 -- 2.26 \\
 \text{8} & 0.122 & 7.8 & 54.2 & 37.7 & 0.3 & 0.0 & 0.39 & 0.308 & 0.17 & 1.25 & 1.20 -- 1.30 \\
 \text{9} & 0.186 & 0.2 & 64.8 & 34.9 & 0.0 & 0.0 & 0.51 & 0.354 & 0.24 & 1.38 & 1.32 -- 1.46 \\
 \text{10} & 0.226 & 3.3 & 31.8 & 63.2 & 1.5 & 0.4 & 0.53 & 0.378 & 0.27 & 2.50 & 2.44 -- 2.56 \\
 \text{11} & 0.186 & 1.9 & 19.7 & 78.3 & 0.1 & 0.0 & 0.41 & 0.354 & 0.27 & 3.46 & 3.44 -- 3.49 \\
 \text{12} & 0.766 & 4.8 & 44.9 & 48.1 & 1.9 & 0.2 & 0.73 & 0.568 & 0.32 & 1.96 & 1.86 -- 2.08 \\
 \text{13} & 0.535 & 1.1 & 22.3 & 70.0 & 4.8 & 1.7 & 0.62 & 0.504 & 0.39 & 3.58 & 3.46 -- 3.70 \\
 \text{14} & 0.063 & 4.3 & 30.6 & 65.0 & 0.0 & 0.0 & 0.33 & 0.246 & 0.18 & 2.42 & 2.30 -- 2.54 \\
 \text{15} & 0.036 & 4.7 & 57.2 & 35.9 & 2.0 & 0.4 & 0.29 & 0.205 & 0.14 & 1.18 & 1.11 -- 1.27 \\
 \text{16} & 0.068 & 2.7 & 43.1 & 54.2 & 0.1 & 0.0 & 0.29 & 0.253 & 0.17 & 2.12 & 2.09 -- 2.15 \\
 \text{17} & 0.182 & 2.9 & 40.2 & 56.8 & 0.1 & 0.0 & 0.45 & 0.351 & 0.22 & 2.04 & 1.94 -- 2.15 \\
 \text{18} & 0.429 & 2.1 & 29.6 & 63.9 & 2.8 & 1.6 & 0.66 & 0.468 & 0.33 & 2.95 & 2.68 -- 3.28 \\
\bottomrule
\end{tabular}
\end{threeparttable}

\label{tab:results}
\end{table*}


The calculations of the steady state temperature profile of non-rotating chondrules showed that the orientation of the particle with respect to direction of illumination has some, but only a small influence on the photophoretic force.
The absolute force varies by only 4 \% on average.
The direction of the force deviates only up to $3^\circ$ on average around the direction of illumination.
More details can be found in Paper I.

In contrast to the findings of Paper I, the first experiments by \citet{Wurm2010} indicated strong variations in the photophoretic force for individual chondrules; however, due to the uncertainties, these first experiments did not allow more specific statements.
The question behind this paper is whether experiments agree with the deduced photophoretic forces for the specific sample of chondrules.
We therefore improved the experimental setup and carried out a new set of experiments at the Bremen drop tower facility.

\section{DROP TOWER EXPERIMENTS}\label{sec:experimentalAspects}

\begin{figure}[h!]
   \centering
   \includegraphics[width=\columnwidth]{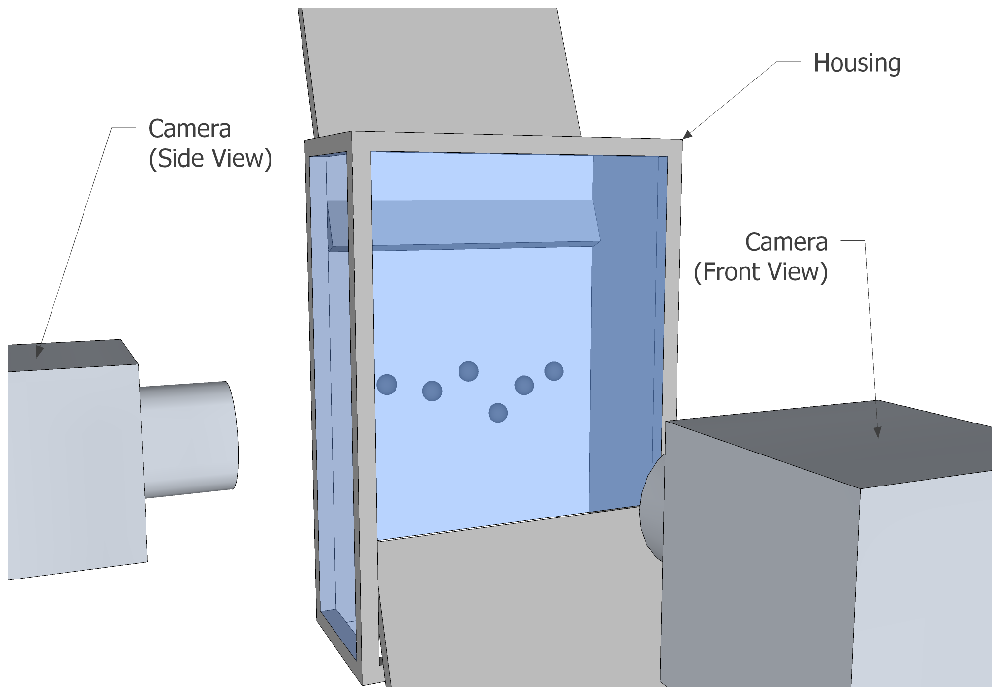}
   \caption{Experimental setup: two cameras observe the chondrules, which are released by two sliders within a small glass housing. The chondrules are illuminated from the top by a laser (not shown here).}
   \label{fig:experiment-closeup}
\end{figure}

The drop tower experiments were carried out in catapult mode, where the experiment is not just dropped from a certain height but is launched upwards. As the experiment is in free ``fall'' as soon as it leaves the launcher, this essentially doubles the microgravity time compared to a drop to about 9 s. The experimental setup is shown in Figure \ref{fig:experiment-closeup}. The chondrules are confined in a glass housing (20 mm* 12 mm* 40 mm) to guarantee that no particles can leave the observable volume. The particle motion is observed with two cameras at 500 frames/s. The cameras are aligned perpendicular to each other, so the three-dimensional particle tracks can be determined. The particles are illuminated with a laser from he top (not shown in Figure \ref{fig:experiment-closeup} for simplicity) with a light intensity of $I = 41.3 \pm 4.5 \; \mathrm{kW/m^2}$. The laser direction is perpendicular to both cameras.

During the launch, the entire experimental apparatus is strongly accelerated.
This also puts tension on the setup that is released after the onset of the microgravity phase of the experiment.
The chondrules are kept in small cavities in a sample mount during this time.
Only 200 ms after the onset of microgravity, the two sliders are pulled outwards to release the chondrules into the middle of the glass housing.
After release, chondrules only move slowly as collisions within the cavities damped most of a particle's motion induced by the launch.
In general, chondrules were released with an arbitrary initial velocity, typically smaller than 2 cm/s and an initial rotation frequency on the order of less than 10 Hz, although the rotation cannot be rigorously quantified.

The glass housing, together with the two sliders of the sample mount, is placed in a vacuum chamber, which is not shown in Figure \ref{fig:experiment-closeup}. All experiments were carried out at pressures between 9 Pa and 50 Pa, where photophoresis is strong enough to lead to a readily detectable acceleration of the chondrules.

\begin{figure}[h!]
   \centering
   \includegraphics[width=\columnwidth]{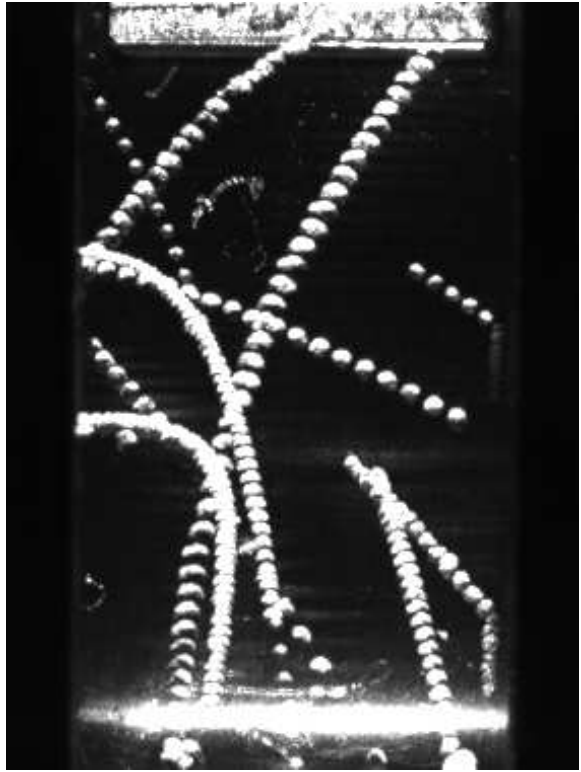}
   \caption{Example for particle trajectories as seen with the front camera visualized as an overlay of several exposures.}
   \label{fig:trajectories}
\end{figure}

The particle trajectories are determined directly from the camera images. The vertical component ($z$) is determined from both camera positions, so the mean value of both data sets is taken. From each data set, one horizontal component of the particle trajectory can be  determined ($x$ component from the front view, $y$ component from the side view). The frame rate of the two cameras directly gives the time steps needed to determine velocity and acceleration of the chondrules. Assuming a constant acceleration, the trajectories are fitted by parabolas ($s = 0.5 \, a_1 \, t^2 + a_2 \, t + a_3$). Figure \ref{fig:trajectories} gives an example for a camera image (front view). The image is an overlay of several exposures to demonstrate how chondrules are accelerated by photophoresis.

To quantify the photophoretic force acting on the chondrules, the mass of each chondrule was determined to better than $\pm 1\%$.

To evaluate the experimental data further, the pressure dependence of photophoresis has to be taken into account. The experiments were not carried out in the free molecular flow regime, but in the transition regime and at varying environmental gas pressure. Therefore, we use Equation \eqref{eq:forceNumerical} to deduce the maximum photophoretic force $F_\mathrm{max}$  \citep{Rohatschek1995}:
\begin{equation}
   F_\mathrm{max} = F \frac{\frac{p_\mathrm{max}}{p} + \frac{p}{p_\mathrm{max}}}{2} \label{eq:Fmax}
\end{equation}
with
\begin{equation}
   p_\mathrm{max} = \sqrt{\frac{3\pi\,\kappa_\mathrm{cr}}{2\alpha}} \, \eta \frac{\overline{c}}{r} \label{eq:pmax}
\end{equation}
which is a pressure-independent measure of the photophoretic force for a given particle.
$\kappa_\mathrm{cr}$ is the thermal creep coefficient, assumed to have a value of 1.14, $\eta$ is the dynamic viscosity, and $\overline{c}$ is denoting the average velocity of the gas molecules.
We assume $\alpha = 0.7$.
The values for total volume, compounds, and radii ($r$) of the chondrules listed in Table \ref{tab:results} are determined from the X-ray tomography results by applying thresholds to the image stacks (see Table 1 in Paper I). For those chondrules, the radius of a sphere with a corresponding volume is taken into account ($r=r_{\text{s}}$; Table \ref{tab:results}).
The detailed procedure is described in Paper I.
For some chondrules, no tomography data exists, therefore these chondrules are not listed in Table \ref{tab:results} as no numerical model exists.
Their radius is then determined by two-dimensional (2D) image analysis.
Images are taken from several positions and the cross section is determined.
The radius is then calculated as the radius of a sphere with the same mean cross section.
\section{EXPERIMENT VERSUS STATIC MODEL}

The photophoretic force for each chondrule was calculated by means of Equation \eqref{eq:forceNumerical} (see Paper I for details).
The modeled forces only vary on the percent level depending on the orientation of the chondrule with respect to the light source.
It has to be noted, though, that the absolute value is associated with a systematic uncertainty. As we do not know the exact thermal conductivities of the individual components, especially of the mesostatis, the force might systematically be too low or too high.
Also, chondrules do not absorb light perfectly, which adds another currently unknown factor.
We estimate this possible offset to a factor of two or three for a given chondrule, but this is a rough guess and we cannot specify this in more detail here.
This is of minor importance here as the variability of the photophoretic force for an individual chondrule is to be discussed.

To compare the calculations to the experiments, all forces were calculated for or scaled to the optimum pressure ($p_\mathrm{max}$) resulting in the maximum force $F_\mathrm{max}$.
The specific light flux density used in the experiments was also used for the
calculations.

\begin{figure}[h!]
   \centering
   \includegraphics[width=\columnwidth]{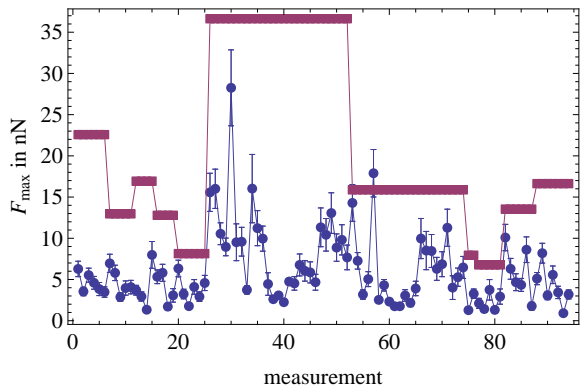}
   \caption{$F_\mathrm{max}$ calculated (red) and measured (blue) for different chondrules. Same/identical model values imply that the same chondrule was measured several times.}
   \label{fig:Fmax}
\end{figure}

\begin{figure}[h!]
   \centering
   \includegraphics[width=\columnwidth]{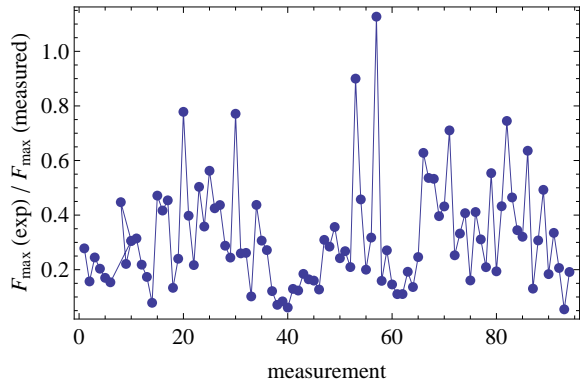}
   \caption{Ratio between experimental and calculated force using the same data set as in Figure \ref{fig:Fmax}.}
   \label{fig:ratios}
\end{figure}

Figure \ref{fig:Fmax} shows the theoretical and experimental forces determined for
the chondrules. The steps mark chondrules that were measured several times,
which always results in the same modeled value but the experimental value varies 
strongly.
The variation is more clearly seen in Figure \ref{fig:ratios}.
Even when considering that the modeled value has some offset, the experimental values are not in agreement with the predictions of Paper I. On average, experimental values are a factor of three smaller than the model predicts and there is up to a factor of 10 scatter in ratios between the experiment and the model. The static equilibrium model in Paper I suggested that such variations are not plausible.

\begin{figure}[tbp]
   \centering
   \includegraphics[width=\columnwidth]{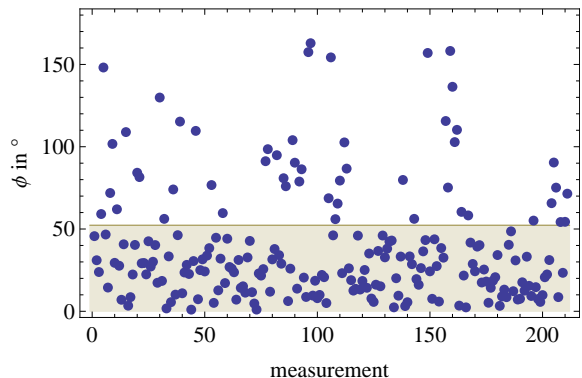}
   \caption{Angles between photophoretic force and the direction of incident light.
      Numerical modeling predicts angles smaller than $52^\circ$ for long duration constant rotation around one well-defined axis.}
   \label{fig:sidewards}
\end{figure}

One more aspect that can be deduced from the experimental data is the ratio between photophoretic force perpendicular to the light source and along the direction of the incident light or the angle between force and illumination. The models in Paper I suggested that this should not be larger than $3.0^\circ\pm 1.5^\circ$ on average; Figure \ref{fig:sidewards} shows the experimental values.

A much stronger variation can also be seen here along with the average forces.
Most data seem to be restricted to angles below about $50^\circ$.

There are two differences remaining between the calculations and the experiments that might induce variations: (1) the particles rotate and (2) there is a time evolution in heating the chondrules.

\section{TIME EVOLUTION MODELS}\label{sec:numericalAspects}

Static equilibrium models are not consistent with the experimental data. To evaluate whether heating and rotation can explain the difference, time-dependent modeling is
required.
We therefore extended the calculations of Paper I and used COMSOL v4.3b to solve the time dependent heat transfer equation with thermal radiation and irradiation:
\begin{equation}
	\boldsymbol{\nabla} \cdot k\boldsymbol{\nabla} T = \rho \, c \, \partial_t T \; . \label{eq:heatEq}
\end{equation}
As before, as a boundary condition, we included cooling through thermal emission in all directions and heating through absorption of illumination from a given direction.
We set emissivity to $1$ , the light flux to $I = 20\,\mathrm{kW/m^2}$ and  $T_\mathrm{gas} = 301\,\mathrm{K}$.

\subsection{Heat up, no rotation}\label{subsec:timeDependentForces}

\begin{figure*}[tbp]
   \centering
   \subfigure[poor conductors]{
      \includegraphics[width=\columnwidth]{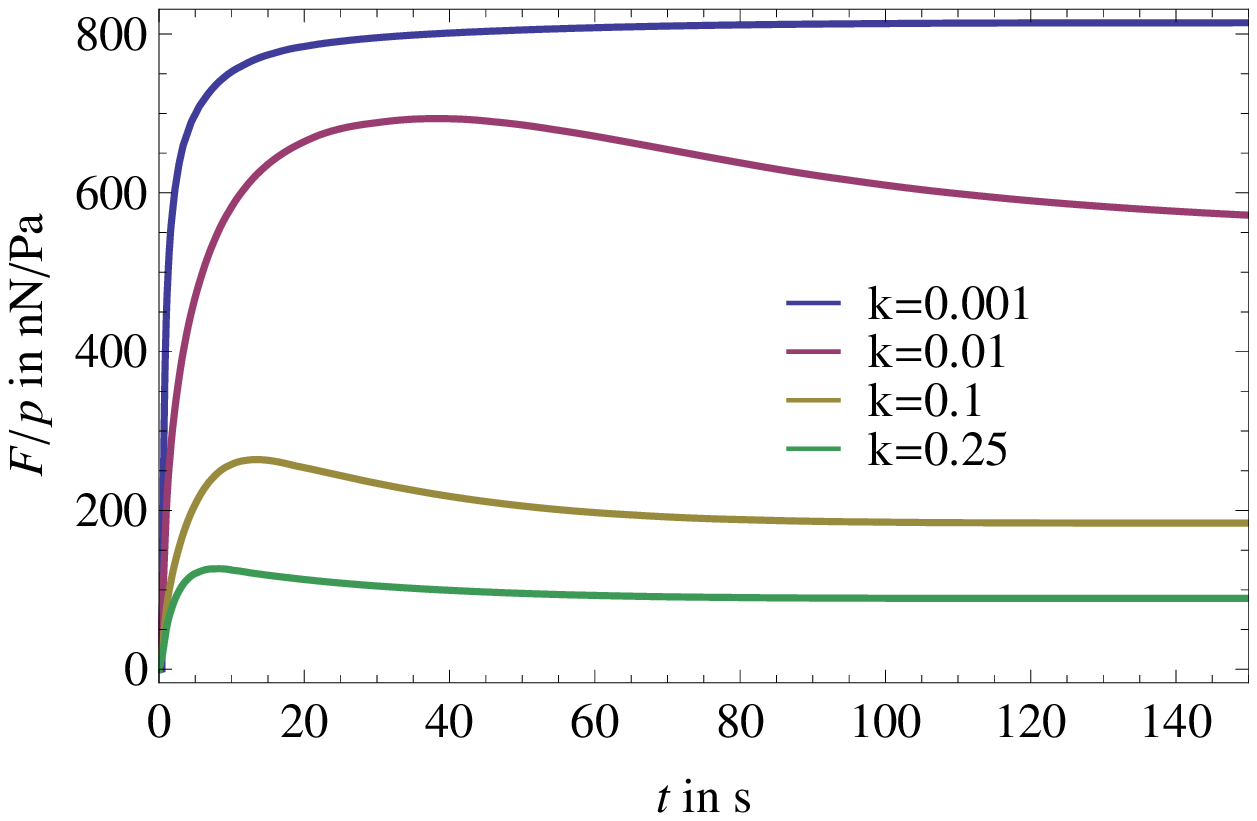}
      }
        ~ 
   \subfigure[good conductors]{
      \includegraphics[width=\columnwidth]{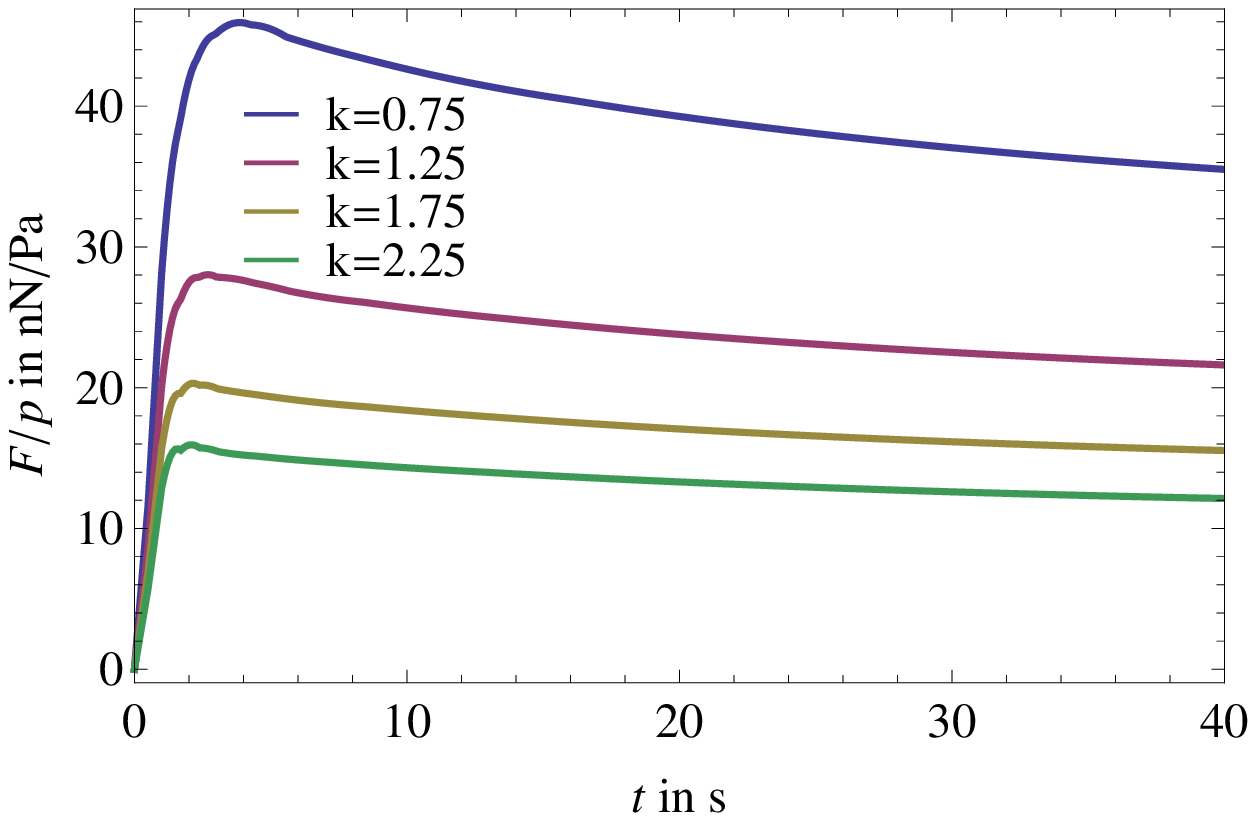}
      }
   \caption{Time evolution of the photophoretic force if the light source is switched on at time $t = 0\,\mathrm{s}$ ($\alpha = 1, r=1\,\mathrm{mm}$); thermal conductivities $k$ in $\mathrm{W/(K\,m)}$.}
   \label{fig:F(t)}
\end{figure*}

For non rotating bodies we calculated the time dependence of the photophoretic force for particles in the following range of thermal conductivities and particle sizes:
\begin{itemize}
   \item $k: 10^{-3}\dots 4\,\mathrm{W/(m\,K)}$ \; \text{and}
   \item $r: 10^{-4}\dots 10^{-2}\,\mathrm{m}$ \; .
\end{itemize}
Typical profiles for poor and good conductors are seen in Figure \ref{fig:F(t)}.
Two timescales appear: a relatively fast rise to a maximum value and a rather slow
decline toward the static equilibrium. We attribute the fast rise to the absorption of
radiation and heating of the illuminated side and the slow decline to heat transfer through the particle that eventually will also increase the temperature at the back side of the particle.
We note that the time dependence of each calculation can be approximated by the following function:
\begin{equation}
   F(t)=\begin{cases}
   a \, \mathrm{sech}\left(\frac{t-b}{c}\right) \left(1-e^{f-t/\tau}\right) + g & k\geq 0.4\,\mathrm{W/(m\, K)}\;,\\
   \left(d-\frac{a}{e^{-t/b}+c}\right) \left(1-e^{f-t/\tau}\right) + g & \text{otherwise.}
   \end{cases} \label{eq:F(t)}
\end{equation}
This shows the exponential decay toward larger times but we currently do not have a closed form for all fit parameters depending on thermal conductivity and particle size. 

\begin{figure}[tbp]
   \centering
   \includegraphics[width=\columnwidth]{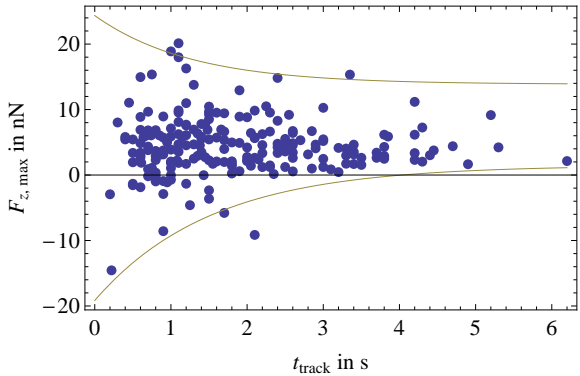}
   \caption{Photophoretic force ($z$ component) over the duration of a trajectory. Solid lines emphasize tendency (to guide the eye; no fit).}
   \label{fig:tracktimes}
\end{figure}

For chondrules ($\approx 1$ $\rm{W/(K\,m)}$) the warming time is about 1--2 s (Figure \ref{fig:F(t)}). Hence, measuring the photophoretic force on chondrules for similar timescales immediately after the light source was switched on should result in smaller forces. If the chondrule was pre-heated (e.g., within the sample holder), large forces should be measured. If a chondrule bounced from a wall, was reoriented, and measured for short timescales, negative $z$ components (toward light source) should also be possible. In total, large variations should occur if photophoretic forces were extrapolated from tracks with short duration. This is consistent with the experiments (Figures \ref{fig:tracktimes} and \ref{fig:sidewards2}). Part of the discrepancy between the experiment and the numerical model can therefore be explained by the time-dependent evolution of the photophoretic force for particles in the experiment.

For longer timescales the variations should become smaller, but it should be noted that only a few observed track times were larger than 3 s. As this corresponds to about two warm-up times, only track times larger than about 3 s should converge on an equilibrium force for non rotating particles. This tendency is visible in Figure \ref{fig:tracktimes}.
However, some factor remains as a variation in the data even if only data restricted to long durations are considered. 
This might be due to particle rotation, which is discussed in the following section.

\subsection{Rotation}\label{sec:particleRotation}

\begin{figure}[tbp]
   \centering
   \includegraphics[width=\columnwidth]{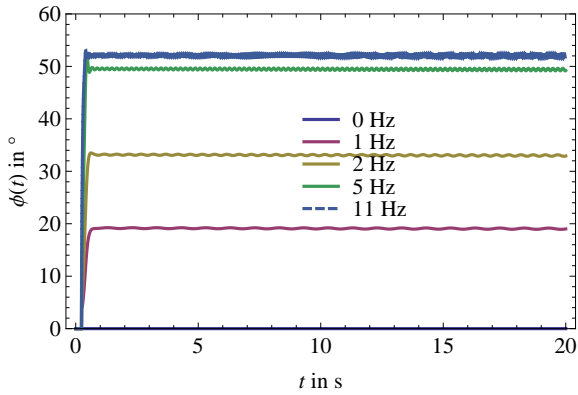}
   \caption{Angle between photophoretic force and incident light.}
   \label{fig:angles}
\end{figure}

\begin{figure}[tbp]
   \centering
   \includegraphics[width=\columnwidth]{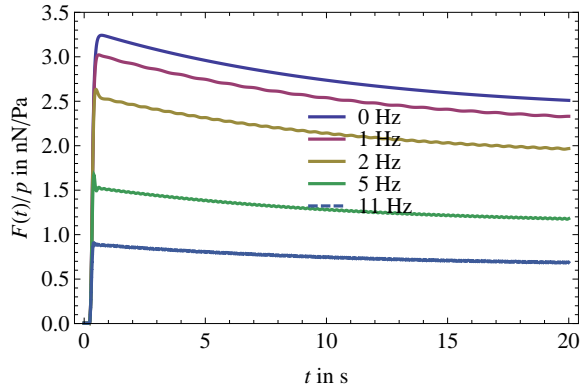}
   \caption{Evolution of the photophoretic force $F(t)$ with time ($\alpha=1$).}
   \label{fig:ft}
\end{figure}

The surface temperature and photophoretic force on a rotating particle was modeled for different rotation frequencies between 0 and 12 Hz. The particle rotation has two consequences. It decreases the absolute force, but it also changes its direction as the warmer part of the surface trails into the shadow. The photophoretic force $\mathbf{F}(t)$ and the irradiation $\mathbf{I}(t)=I*\mathbf{e}_I$ enclose an angle
\begin{equation}
   \phi(t)= \angle (\mathbf{F}(t), \mathbf{I}(t)) \; ,
\end{equation}
which describes a phase lag between the incident laser and heat transfer reacting to it (Figure \ref{fig:angles}).
Due to computational limitations, we constrained the calculations to study the influence of rotation  for one fixed particle size $r=0.568\,\mathrm{mm}$ and one fixed thermal conductivity $1.96\,\mathrm{W/(m\cdot K)}$. These values correspond to those deduced for one of the chondrules in Paper I (also see Table \ref{tab:results}). Again, we can consider the absolute force depending on rotation frequency and the angle between force and
illumination.
The time evolution of the absolute value of the force shows a similar behavior as in the case of a non-rotating particle (Figure \ref{fig:ft}).

\begin{figure}[tbp]
   \centering
   \includegraphics[width=\columnwidth]{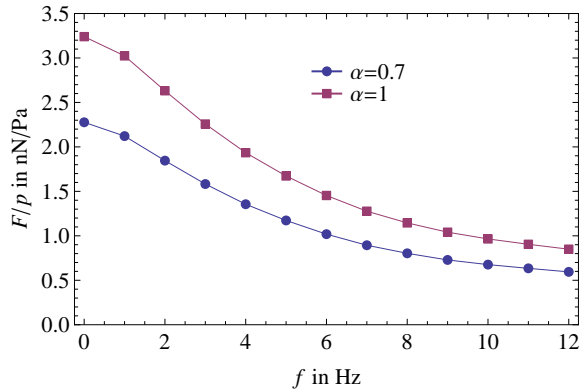}
   \caption{Maximum photophoretic force $F$ ($\max_t F(t,\omega)$) over rotation frequency $f$.}
   \label{fig:fmaxen}
\end{figure}

There is a rapid increase, a pronounced maximum after a second, and a slow decrease to about 70\% of the maximum value with a decay time of about 10 s. Obviously the rotation does not change the principle mechanisms of heating and cooling. However, the absolute values decrease with increasing rotation frequency. This is more clearly shown in Figure \ref{fig:fmaxen}, which shows the maximum force depending on rotation frequency.
The total photophoretic force varies by a factor of three to four in the studied frequency range. Rotation can therefore strongly reduce the absolute value of the photophoretic force. In agreement with the experiments, this can explain the strong variations seen in the experiments, even for chondrules with the longest track times larger than 3 s. 
The phase lag between force and illumination can be seen in Figure \ref{fig:angles}.

\begin{figure}[tbp]
   \centering
   \includegraphics[width=\columnwidth]{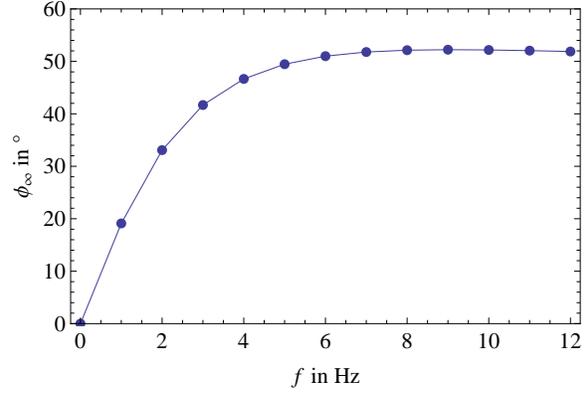}
   \caption{Equilibrium angle between photophoretic force and incident light.}
   \label{fig:anglelimits}
\end{figure}

The misalignment increases as the frequency increases. However, at high 
frequencies the angle reaches a maximum value of about $52^\circ$ as seen in Figure \ref{fig:anglelimits}.

\begin{figure}[tbp]
   \centering
   \includegraphics[width=\columnwidth]{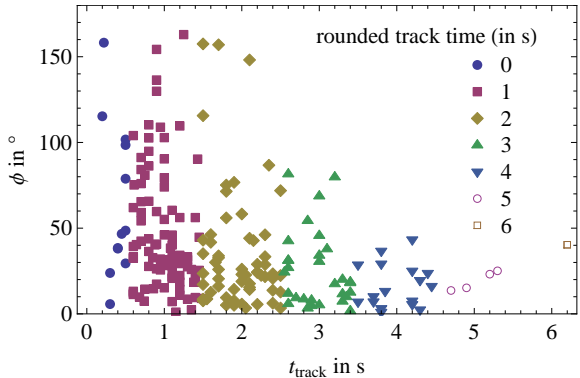}
   \caption{Angles between photophoretic force and the direction of incident light for different tracking times.}
   \label{fig:sidewards2}
\end{figure}

In general, this fits to a strong clustering of the experimental values below $50^\circ$ (Figure \ref{fig:sidewards}). However, this simple model excludes angles larger than $52^\circ$, which are observed in the experimental data. In our experiments, the chondrules have initial velocities and collide with the walls. As stated above, a pre-heated chondrule allowed to readjust after it was released or after it collided with a wall can show negative photophoretic forces until the temperature gradient is equilibrated. Therefore, large angles should be possible for a few heat-up timescales or below about 3 s. This is consistent with the experimental data. Figure \ref{fig:sidewards2} connects the angles with the respective duration of the measured tracks (tracking time).

\begin{figure}[tbp]
   \centering
   \includegraphics[width=\columnwidth]{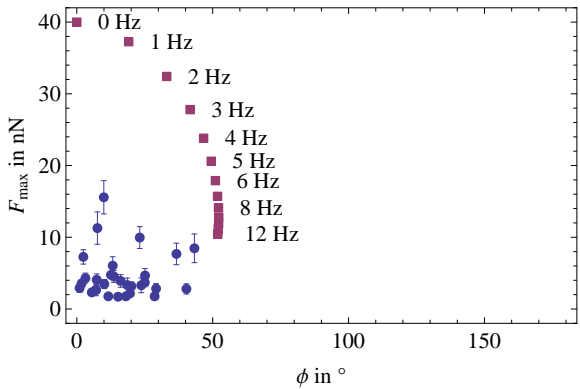}
   \caption{Exp. scattering angles $\phi$ vs. $F_{\max}$ for tracking times $t_\text{track} > 3.2 \,\mathrm{s}$ (blue circles). All values are enveloped by the limits predicted by the rotation modeling ($\alpha=0.7$) (red squares).}
   \label{fig:anglesVsForce}
\end{figure}

The modeling shows that the photophoretic force and the angle 
both depend on the rotation frequency. Combined, this gives a relation between force
and angle. In general, a reduction in force should be correlated to an increase in angle, approaching the equilibrium value eventually. 
This relation and the experimental values are seen in Figure \ref{fig:anglesVsForce}.

Although the experimental force and angle values should fall on the modeled line for ideally rotating particles, they do not and are constrained to smaller values (Figure \ref{fig:anglesVsForce}). This is to be expected. Rotation decreases the photophoretic force, but the angle is strongly related to the direction of rotation. Rotation in the experiments is not fixed about one axis. Photophoresis on the non-perfectly spherical particles results in torque and particle precession. An indication that the rotation of a chondrule can change on sub second timescales was already published in \citet{Wurm2010}. Even in a torque-free case, a nutation can change the rotation axis systematically. The effect will be that the photophoretic force decreases due to rotation, but the measured angle is no longer correlated to this decrease and can reach values between zero and the maximum.

\subsection{Photophoretic Yarkovsky analog}\label{subsec:yarkovskiAnalogue}

The observed behavior of the photophoretic force on the rotation of the particle is not unexpected.
In the extreme case, for a very fast rotating particle, all sides should be equally heated and the absolute photophoretic force has to decrease.
For slower rotations, the warm heated part of a particle trails somewhat into the shadowed side and vice versa, which implies a sideward component of the photophoretic force.

This is closely related to the Yarkovsky effect, which describes the change of the orbit of a rotating asteroid due to radiation pressure \citep{Rubincam1995}.
Such changes have indeed been observed \citep{Chesley2003}.
The ideas are similar to our considerations given above.
An illuminated asteroid has a warm day and a cooler night side.
If the asteroid rotates, the warm side can trail into the night side and the surface temperature is no longer symmetric to the direction of illumination.
This is exactly the same as for the rotating chondrules. The sideward photophoretic force originates as gas molecules at the warmer surface are re-ejected with larger momentum and the particle has to balance this. Radiation does the same.
At the warmer surface parts, more radiation is emitted due to the Stefan-Boltzmann law ($\propto \sigma \, T^4$).
As the photons also carry momentum, the asteroid has to balance this.

Essentially, it is only a substitution between gas molecules and photons that gives the difference in absolute magnitude, but otherwise the mechanisms are the same.
Photophoresis of a rotating chondrule is therefore closely approximating a Yarkovsky-like effect.
The experiments together with the heat transfer simulations can be regarded as experimental proof of the effect.
A study of photophoretic interaction with non-symmetric particles should also allow future model experiments of, e.g., the YORP effect, where rotational momentum due to radiation pressure leads to a change of the rotation of asteroids. As outlined above, such changes in rotation are clearly present in the current experimental data.

\begin{figure}[tbp]
   \centering
   \includegraphics[width=\columnwidth]{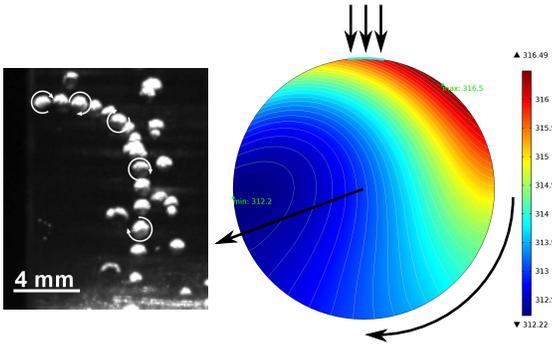}
   \caption{Rotation of a particle superimposed on its trajectory (left); temperature distribution on a rotating chondrule and direction of photophoretic force (right).}
   \label{fig:rotation}
\end{figure}

Unfortunately, our experiments were not optimized to study rotation, and the almost spherical shape of the particles and spatial resolution limits do not allow a quantitative reconstruction. In most cases, the particle rotation also includes a nutation or precession and is therefore not fixed to one axis.
Nevertheless, even without detailed analysis, the general statement can be made that the particles rotate in the experiments with frequencies of several Hz.
Some experimental data imply that the direction of rotation is linked to the sideward motion of the particle (Figure \ref{fig:rotation}).
As illustrated, the particle moves perpendicular to the direction of light as it rotates in a direction fitting to a photophoretic Yarkovsky effect.
The simulations give the same direction of motion.

\section{TRANSPORT OF CHONDRULES IN PROTOPLANETARY DISKS}
Steady state calculations of heat transfer through chondrules with composition determined by tomography showed that the photophoretic force only slightly depends on the particle orientation. On average, the absolute force for an individual chondrule varies by only 4\% and is directed along the direction of light within $3^\circ$ (Paper I). This allows a simple yet accurate analytic calculation of the average photophoretic force acting on a chondrule. The chondrule can be described as a spherical particle of equivalent volume with an average thermal conductivity determined by the volume ratio of the two dominant components, coarse-grained silicates and mesostasis (Paper I).

However, the measured forces during short duration microgravity experiments are in stark contrast to the static models. They are smaller by a factor of three. The forces for individual chondrules, which were measured several times each, showed large variations up to a factor of 10. The deviations from the direction of illumination included almost all angles (this paper).

This seeming contradiction between model and experiment can be explained by extending the static model to a time evolution model and including the rotation of the chondrules. The results of such calculations indicate that the difference between static model and experiment is due to the warm up time and the heat redistribution by the rotation.  A particle spinning around an axis perpendicular to the direction of light leads to a reduction of the photophoretic force and changes its direction (this paper).

What does this mean for the motion of illuminated chondrules in the early solar nebula? The answer is tied to particle spin in a protoplanetary disk. If particles spin randomly, the photophoretic force can still be significant (it will never vanish entirely), but the large deviation would not allow for any sorting due to composition and size. So more fundamentally, the essence of this work is related to the question ``if'' and ``how'' particles really spin in protoplanetary disks. There has been much debate about this. The answer has different aspects.
\begin{itemize}
   \item Random rotation: random rotation occurs mostly after collisions. Between collisions the rotations are damped though on a gas--grain friction time. For a millimeter particle at $10^{-5}$ bar, this timescale is in minutes. After a few minutes, any random rotation has ceased. As collision times are much longer, there is no random motion of chondrules in protoplanetary disks of any significance for photophoresis.
   \item Forced rotation: torques induced by external force fields can lead to particle rotation. Torques on a non spherical particle are induced by photophoresis itself. It has to be kept in mind, however, that the particles move within an ambient gas and feel gas friction. In such a damped system, forced rotations have two effects. First they lead to an alignment of the rotation axis with direction of illumination. For photophoresis, this has been shown by \citet{Eymeren2012}. After this alignment, only rotation around the direction of illumination remains. This does not change the front and back sides, though, and the photophoretic strength is the one calculated for the static case. The small differences between the force and direction of illumination together with the remaining rotation simply results in helical motion around the direction of the light. This has recently been shown by \citet{Kuepper2014b} for small basalt grains.
\end{itemize}
We conclude that on the timescales of protoplanetary disks, particle spin is not important for chondrule motion. Photophoretic forces are well constrained and can be calculated. They depend on particle size and composition. The variations for individual chondrules are small and do not strongly depend on orientation. Therefore, size or compositional sorting remains possible by photophoresis.

\section{CONCLUSIONS}
The goal behind this paper and Paper I was to provide an accurate description of the photophoretic motion of chondrules in optical thin parts of the early solar nebula. It combined analytical, numerical, and experimental techniques. Particle rotation was shown to reduce the photophoretic force on timescales of seconds in the drop tower experiments when compared to static models. However, random particle spin in protoplanetary disks does not occur on long timescales due to gas--grain friction. Only a well-constrained forced rotation can be sustained, but with a rotation axis aligned to the illumination. This has no effect on photophoresis other than producing helical trajectories.

We find that the results of Paper I are sound. The particle motion can be calculated to high accuracy using the given equations (e.g., Equation \eqref{eq:forceNumerical} in this paper) and particles might be sorted due to size or composition.
\section{ACKNOWLEDGMENTS}
Part of this work is funded by the Deutsche Forschungsgemeinschaft within the priority program SPP 1385 ``The first ten million years, a materials approach''. We very much appreciate that access to the drop tower has been granted by the ESA (ID 9282).

J.M.F. thanks the Fund for Astrophysical Research for assistance in the acquisition of computer equipment used for portions of this study and NASA’s Origin of Solar Systems (OSS) Program grant NNX10AH336 (Co-I J.M.F.) for additional support. Portions of this work were performed at GeoSoilEnviroCARS (Sector 13), Advanced Photon Source (APS), Argonne National Laboratory. GeoSoilEnviroCARS is supported by the National Science Foundation--Earth Sciences (EAR-1128799) and Department of Energy--Geosciences (DE-FG02-94ER14466). Use of the Advanced Photon Source was supported by the U. S. Department of Energy, Office of Science, Office of Basic Energy Sciences, under Contract No. DE-AC02-06CH11357.
We thank the reviewer for his helpful comments.

\appendix
\bibliographystyle{apalike} 
\bibliography{Referenzen} 

\begin{thebibliography}{}

\bibitem[{Beresnev} et~al., 1993]{Beresnev1993}
{Beresnev}, S., {Chernyak}, V., and {Fomyagin}, G. (1993).
\newblock Photophoresis of a spherical particle in a rarefied gas.
\newblock {\em Physics of Fluids}, 5:2043--2052.

\bibitem[{Cheremisin} et~al., 2005]{Cheremsin2005}
{Cheremisin}, A.~A., {Vassilyev}, Y.~V., and {Horvath}, H. (2005).
\newblock Gravito-photophoresis and aerosol stratification in the atmosphere.
\newblock {\em Journal of aerosol science}, 36(11):1277--1299.

\bibitem[{Chesley} et~al., 2003]{Chesley2003}
{Chesley}, S.~R., {Ostro}, S.~J., {Vokrouhlick{\'y}}, D., {{\v C}apek}, D.,
  {Giorgini}, J.~D., {Nolan}, M.~C., {Margot}, J.-L., {Hine}, A.~A., {Benner},
  L.~A.~M., and {Chamberlin}, A.~B. (2003).
\newblock {Direct Detection of the Yarkovsky Effect by Radar Ranging to
  Asteroid 6489 Golevka}.
\newblock {\em Science}, 302:1739--1742.

\bibitem[{de Beule} et~al., 2013]{Beule2013}
{de Beule}, C., {Kelling}, T., {Wurm}, G., {Teiser}, J., and {Jankowski}, T.
  (2013).
\newblock {From Planetesimals to Dust: Low-gravity Experiments on Recycling
  Solids at the Inner Edges of Protoplanetary Disks}.
\newblock {\em \apj}, 763:11.

\bibitem[{Ehrenhaft}, 1918]{Ehrenhaft1918}
{Ehrenhaft}, F. (1918).
\newblock {Die Photophorese}.
\newblock {\em {Annalen der Physik}}, 361(10):81--132.

\bibitem[{Haack} and {Wurm}, 2007]{Haack2007}
{Haack}, H. and {Wurm}, G. (2007).
\newblock Life on the edge - formation of cais and chondrules at the inner edge
  of the dust disk.
\newblock volume~42, page 5157.

\bibitem[{Herrmann} and {Krivov}, 2007]{Herrmann2007}
{Herrmann}, F. and {Krivov}, A.~V. (2007).
\newblock Effects of photophoresis on the evolution of transitional
  circumstellar disks.
\newblock {\em \aap}, 476:829--839.

\bibitem[{Kelling} and {Wurm}, 2009]{Kelling2009}
{Kelling}, T. and {Wurm}, G. (2009).
\newblock {Self-Sustained Levitation of Dust Aggregate Ensembles by
  Temperature-Gradient-Induced Overpressures}.
\newblock {\em \prl}, 103:215502--1--215502--4.

\bibitem[{Kelling} and {Wurm}, 2011]{Kelling2011}
{Kelling}, T. and {Wurm}, G. (2011).
\newblock {A Mechanism to Produce the Small Dust Observed in Protoplanetary
  Disks}.
\newblock {\em \apj}, 733:120--125.

\bibitem[{Kelling} and {Wurm}, 2013]{Kelling2013}
{Kelling}, T. and {Wurm}, G. (2013).
\newblock {Accretion through the Inner Edges of Protoplanetary Disks by a Giant
  Solid State Pump}.
\newblock {\em \apjl}, 774:L1.

\bibitem[{Krauss} and {Wurm}, 2005]{Krauss2005}
{Krauss}, O. and {Wurm}, G. (2005).
\newblock Photophoresis and the pile-up of dust in young circumstellar disks.
\newblock {\em \apj}, 630:1088--1092.

\bibitem[{Krauss} et~al., 2007]{Krauss2007}
{Krauss}, O., {Wurm}, G., {Mousis}, O., {Petit}, J.-M., {Horner}, J., and
  {Alibert}, Y. (2007).
\newblock {The photophoretic sweeping of dust in transient protoplanetary
  disks}.
\newblock {\em \aap}, 462:977--987.

\bibitem[{Kuepper} et~al., 2014]{Kuepper2014b}
{Kuepper}, M., de~{Beule}, C., {Wurm}, G., {Matthews}, L.~S., {Kimery}, J.~S.,
  and {Hyde}, T.~W. (2014).
\newblock {Photophoresis on polydisperse basalt microparticles under
  microgravity}.
\newblock {\em Journal of Aerosol Science}, 76:126--137.

\bibitem[{Loesche} and {Wurm}, 2012]{Loesche2012}
{Loesche}, C. and {Wurm}, G. (2012).
\newblock {Thermal and photophoretic properties of dust mantled chondrules and
  sorting in the solar nebula}.
\newblock {\em \aap}, 545:A36.

\bibitem[{Loesche} et~al., 2013]{Loesche2013}
{Loesche}, C., {Wurm}, G., {Teiser}, J., {Friedrich}, J.~M., and {Bischoff}, A.
  (2013).
\newblock {Photophoretic Strength on Chondrules. 1. Modeling}.
\newblock {\em \apj}, 778(2):101.

\bibitem[{Moudens} et~al., 2011]{Moudens2011}
{Moudens}, A., {Mousis}, O., {Petit}, J.-M., {Wurm}, G., {Cordier}, D., and
  {Charnoz}, S. (2011).
\newblock The role of photophoresis in the radial transport of hot minerals in
  the solar nebula.
\newblock In {\em Lunar and Planetary Institute Science Conference Abstracts},
  volume~42 of {\em Lunar and Planetary Inst. Technical Report}, page 1409.

\bibitem[{Mousis} et~al., 2007]{Mousis2007}
{Mousis}, O., {Petit}, J.-M., {Wurm}, G., {Krauss}, O., {Alibert}, Y., and
  {Horner}, J. (2007).
\newblock Photophoresis as a source of hot minerals in comets.
\newblock {\em \aap}, 466:L9--L12.

\bibitem[Rohatschek, 1956a]{Rohatschek1956b}
Rohatschek, H. (1956a).
\newblock {\"U}ber die kr{\"a}fte der reinen photophorese und der
  gravitophotophorese (on the forces of pure and gravito-photophoresis).
\newblock {\em Acta Physica Austriaca}, 10:267--286.

\bibitem[Rohatschek, 1956b]{Rohatschek1956a}
Rohatschek, H. (1956b).
\newblock Zur theorie der gravitophotophorese.
\newblock {\em Acta Physica Austriaca}, 10:227--238.

\bibitem[{Rohatschek}, 1995]{Rohatschek1995}
{Rohatschek}, H. (1995).
\newblock Semi-empirical model of photophoretic forces for the entire range of
  pressures.
\newblock {\em Journal of Aerosol Science}, 26(5):717--734.

\bibitem[{Rubincam}, 1995]{Rubincam1995}
{Rubincam}, D.~P. (1995).
\newblock {Asteroid orbit evolution due to thermal drag}.
\newblock {\em \jgr}, 100:1585--1594.

\bibitem[{Steinbach} et~al., 2004]{Steinbach2004}
{Steinbach}, J., {Blum}, J., and {Krause}, M. (2004).
\newblock Development of an optical trap for microparticle clouds in dilute
  gases.
\newblock {\em The European Physical Journal E: Soft Matter and Biological
  Physics}, 15(3):287--291.

\bibitem[{Takeuchi} and {Krauss}, 2008]{Takeuchi2008}
{Takeuchi}, T. and {Krauss}, O. (2008).
\newblock Photophoretic structuring of circumstellar dust disks.
\newblock {\em \apj}, 677:1309--1323.

\bibitem[van {Eymeren} and {Wurm}, 2012]{Eymeren2012}
van {Eymeren}, J. and {Wurm}, G. (2012).
\newblock {The implications of particle rotation on the effect of
  photophoresis}.
\newblock {\em \mnras}, 420:183--186.

\bibitem[von {Borstel} and {Blum}, 2012]{Borstel2012}
von {Borstel}, I. and {Blum}, J. (2012).
\newblock {Photophoresis of dust aggregates in protoplanetary disks}.
\newblock {\em \aap}, 548:A96.

\bibitem[{Wurm}, 2007]{Wurm2007}
{Wurm}, G. (2007).
\newblock {Light-induced disassembly of dusty bodies in inner protoplanetary
  discs: implications for the formation of planets}.
\newblock {\em \mnras}, 380:683--690.

\bibitem[{Wurm} and {Haack}, 2009]{Wurm2009}
{Wurm}, G. and {Haack}, H. (2009).
\newblock {Outward transport of CAIs during FU-Orionis events}.
\newblock {\em Meteoritics and Planetary Science}, 44:689--699.

\bibitem[{Wurm} and {Krauss}, 2006]{Wurm2006}
{Wurm}, G. and {Krauss}, O. (2006).
\newblock Concentration and sorting of chondrules and cais in the late solar
  nebula.
\newblock {\em \icarus}, 180:487--495.

\bibitem[{Wurm} et~al., 2010]{Wurm2010}
{Wurm}, G., {Teiser}, J., {Bischoff}, A., {Haack}, H., and {Roszjar}, J.
  (2010).
\newblock {Experiments on the photophoretic motion of chondrules and dust
  aggregates{\mdash}Indications for the transport of matter in protoplanetary
  disks}.
\newblock {\em \icarus}, 208:482--491.

\end{thebibliography}

\end{document}